\documentclass[conference]{IEEEtran}
\IEEEoverridecommandlockouts

\usepackage{cite}
\usepackage{amsmath,amssymb,amsfonts}
\usepackage{arydshln}
\usepackage{booktabs}
\usepackage{algorithmic}
\usepackage{graphicx}
\usepackage{textcomp}
\usepackage{xcolor}
\usepackage{multirow}
\usepackage{hyperref}
\usepackage{subfigure}
\usepackage{pifont}       
\usepackage{bbding}       
\usepackage{fontawesome}
\usepackage{balance}

\def\BibTeX{{\rm B\kern-.05em{\sc i\kern-.025em b}\kern-.08em
    T\kern-.1667em\lower.7ex\hbox{E}\kern-.125emX}}
\begin{document}

\title{3DGCQA: A Quality Assessment Database for 3D AI-Generated Contents}
%

\author{
    \IEEEauthorblockN{Yingjie Zhou, Zicheng Zhang, Farong Wen, Jun Jia, Yanwei Jiang, Xiaohong Liu, Xiongkuo Min, Guangtao Zhai}
    \IEEEauthorblockA{Institute of Image Communication and Network Engineering, Shanghai Jiao Tong University, China}
    \IEEEauthorblockA{zyj2000@sjtu.edu.cn}
}

\maketitle

\begin{abstract}
Although 3D generated content (3DGC) offers advantages in reducing production costs and accelerating design timelines, its quality often falls short when compared to 3D professionally generated content. Common quality issues frequently affect 3DGC, highlighting the importance of timely and effective quality assessment. Such evaluations not only ensure a higher standard of 3DGCs for end-users but also provide critical insights for advancing generative technologies. To address existing gaps in this domain, this paper introduces a novel 3DGC quality assessment dataset, 3DGCQA, built using 7 representative Text-to-3D generation methods. During the dataset's construction, 50 fixed prompts are utilized to generate contents across all methods, resulting in the creation of 313 textured meshes that constitute the 3DGCQA dataset. The visualization intuitively reveals the presence of 6 common distortion categories in the generated 3DGCs. To further explore the quality of the 3DGCs, subjective quality assessment is conducted by evaluators, whose ratings reveal significant variation in quality across different generation methods. Additionally, several objective quality assessment algorithms are tested on the 3DGCQA dataset. The results expose limitations in the performance of existing algorithms and underscore the need for developing more specialized quality assessment methods. To provide a valuable resource for future research and development in 3D content generation and quality assessment, the dataset has been open-sourced in \href{https://github.com/zyj-2000/3DGCQA}{https://github.com/zyj-2000/3DGCQA}.
\end{abstract}


\section{Introduction}
\begin{figure}[!t]
    \vspace{-0cm}
    \centering
    \includegraphics[width =\linewidth]{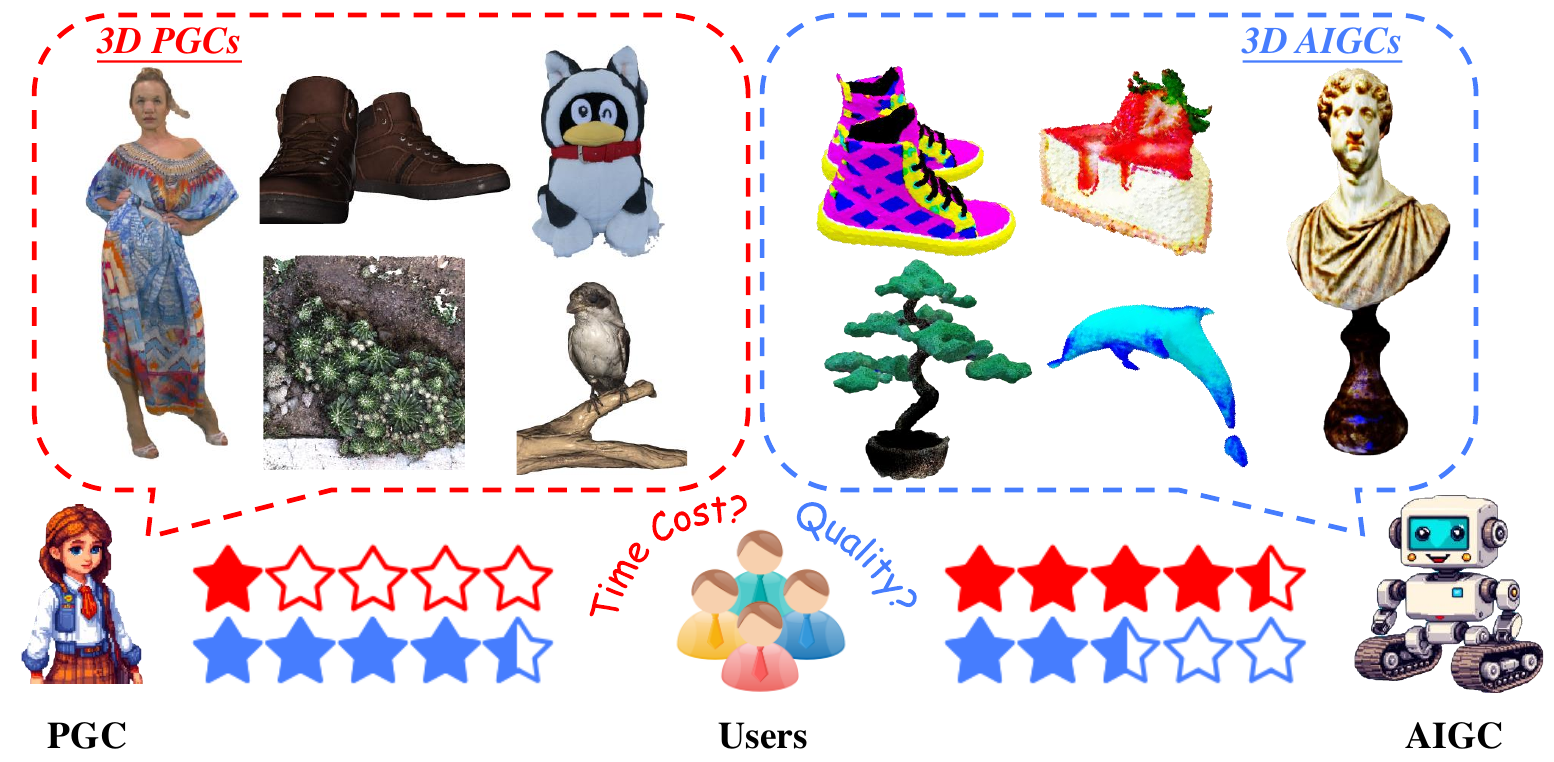}
    \vspace{-0.8cm}
    \caption{Differences in time cost and quality between 3D PGCs and AIGCs.}
    \label{fig:teaser}
    \vspace{-0.4cm}
\end{figure}

In recent years, artificial intelligence generated content (AIGC) has opened up new possibilities for various aspects of life and artistic creation, finding widespread application across industries such as business, film, television, and entertainment \cite{zhou2023implementation}. Generative technology now possesses the capability to autonomously generate audio, images, videos, and even 3D content, thereby lowering the barriers to entry in these fields and significantly enhancing the efficiency of creative professionals. With the rising prominence of virtual reality and the metaverse, 3D AIGC has garnered increased attention and is regarded as both a highly promising and technically challenging area \cite{zhou2023perceptual}. Despite these advancements, as illustrated in Fig.~\ref{fig:teaser}, the quality of most current 3D AIGCs lags behind that of 3D professionally generated contents (PGCs), and often fails to meet users' expectations. This discrepancy can be attributed to the relatively early stage of development in 3D generation technology, as well as the absence of standardized evaluation criteria, which hinders timely and accurate assessments of 3D AIGC quality, thus limiting the ability to provide effective feedback and guidance for improvement.

To address the challenge of assessing 3D Generated Content (3DGC), a publicly available quality assessment dataset, 3DGCQA, has been developed. This dataset is constructed using seven representative generative algorithms, applied to 50 distinct prompts, resulting in the successful generation of 313 3DGCs. Additionally, the generation time for each 3DGC is recorded to evaluate the performance of the different generative algorithms. Volunteers are then invited to assess the quality of the 3DGCs within the dataset. The results from the subjective evaluations reveal significant quality variations and differences in consistency with the prompts across the different generative algorithms. These findings underscore the critical importance of quality assessment for 3DGC. Furthermore, this study tests several commonly used objective quality assessment algorithms to benchmark their performance on 3DGCQA dataset. The experimental results demonstrate the limitations of existing objective assessment methods and highlight the need for the development of more targeted assessment algorithms.

\begin{figure*}[!t]
    \vspace{-0cm}
    \centering
    \includegraphics[width =1\linewidth]{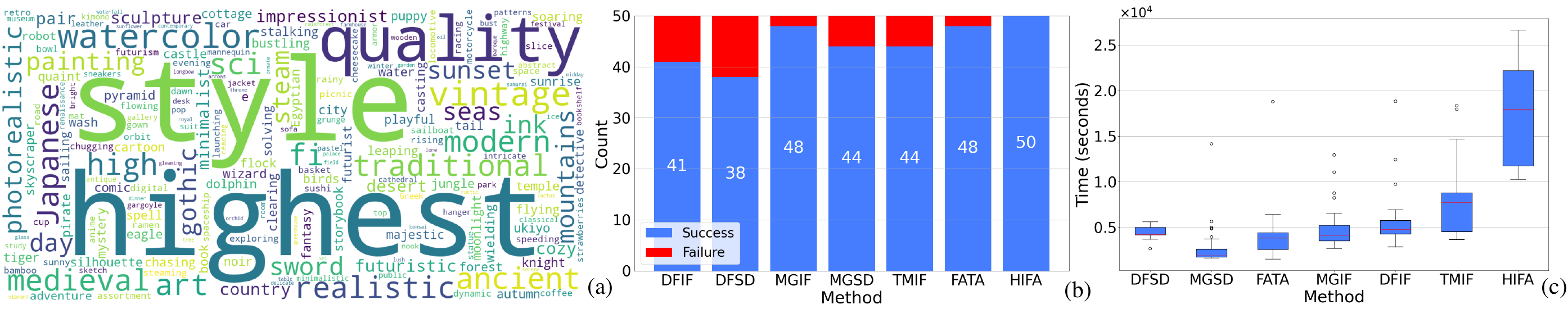}
    \vspace{-0.7cm}
    \caption{Details of the selected prompts and methods. (a) Word clouds consisting of prompts. (b) Number of 3DGCs successfully generated by each method. (c) Generation time of each method on RTX A6000 GPU.}
    \label{fig:prep}
    \vspace{-0.5cm}
\end{figure*}

\section{Related Work}
\subsection{3D AIGC}
3D content plays a crucial role in computer graphics, with two of the most prominent 3D data structures being point clouds and polygonal meshes. Point clouds represent the geometry and color of 3D content through the position and color of a discrete set of points, whereas polygonal meshes separate the geometric and color information. Specifically, meshes describe the structural aspects of 3D content through vertices, edges, and normal vectors, while texture mapping is employed to capture the color information.  In recent years, a new data structure known as 3D Gaussian Splatting (3DGS) \cite{kerbl20233d} has emerged, gaining significant attention within the field. Building on the advancements in these data structures and generative models, numerous 3D generation methods have been developed, leading to promising progress. Based on the type of input guidance, these AI-driven generation methods can be categorized into two main groups: Image-to-3D \cite{qian2023magic123,deng2023nerdi,melas2023realfusion,tang2023make,yu2023hifi} and Text-to-3D \cite{zhu2023hifa,chen2023fantasia3d,lin2023magic3d}. However, most Text-to-3D methods still rely on transforming text prompts into images via Text-to-Image (T2I) models before generating 3D content. Additionally, Text-to-3D methods lack inherent shape information, making their application both more versatile and more challenging. This paper focuses primarily on Text-to-3D methods and aims to offer valuable insights for the continued development and refinement of these technologies.

\subsection{3D Quality Assessment}
3D Quality Assessment (3DQA) can be broadly divided into two categories: Point Cloud Quality Assessment (PCQA) and Mesh Quality Assessment (MQA). Both areas currently have sufficient datasets \cite{lspcqa,basics,sjtupcqa,wpc,dhhqa,ddhqa,h3d,6gqa,zhou2024subjective} to support further research, covering a wide range of content including objects, plants, humans, sculptures, and large-scale scenes. Based on these established datasets, several representative 3DQA algorithms have been developed. Generally, these methods can be classified into model-based \cite{cignoni1998metro,Mekuria2016EvaluationCF} and projection-based \cite{zhou2023no,zhang2023gms,zhang2023eep,zhang2024reduced,chen2023no,zhang2023geometry,zhang2023simple} approaches. Model-based methods assess the geometrical structure of 3D models but often involve high computational costs. In contrast, projection-based methods reduce computational overhead by projecting 3D content from fixed or multiple viewpoints, thereby transforming the 3D quality assessment problem into a 2D quality assessment problem \cite{thqa,zhang2023subjective,sun2024dual}. Recently, the rise of Large Multimodal Model (LMM) has introduced new perspectives for 3DQA \cite{zhang2024quality}, leading to developments such as LMM-PCQA \cite{zhang2024lmm} for point clouds and large-scale model-assisted assessment method \cite{zhou2024subjective} for meshes. Despite the progress in 3DQA, current datasets and evaluation algorithms primarily focus on captured 3D content, while 3DGC has received limited attention. The 3DGQA dataset introduced in this paper aims to address this gap in generative 3DQA and provide valuable reference points for improving existing generative techniques.

\section{Database Construction}

\begin{table}[!tp]\small
    \centering
    \renewcommand\arraystretch{1}
    \caption{Details of selected generation methods. }
    \resizebox{0.95\linewidth}{!}{\begin{tabular}{c|c|c|c}
    \toprule
          Label &Methods & Year & T2I Model \\ \hline
        DFIF & DreamFusion \cite{poole2022dreamfusion}& 2022& DeepFloyd IF \cite{saharia2022photorealistic}\\  
        DFSD & DreamFusion \cite{poole2022dreamfusion}& 2022& Stable Diffusion \cite{rombach2022high}\\        
        MGIF & Magic3D \cite{lin2023magic3d}& 2023& DeepFloyd IF \cite{saharia2022photorealistic}\\  
        MGSD & Magic3D \cite{lin2023magic3d}& 2023& Stable Diffusion \cite{rombach2022high}\\
        TMIF & TextMesh \cite{tsalicoglou2024textmesh}& 2024&DeepFloyd IF \cite{saharia2022photorealistic}\\
        FATA & Fantasia3D \cite{chen2023fantasia3d}& 2023& Stable Diffusion \cite{rombach2022high}\\
        HIFA & HiFA \cite{zhu2023hifa}& 2023& Stable Diffusion \cite{rombach2022high}\\

    \bottomrule
    \end{tabular}}
    \label{tab:methods}
    \vspace{-0.3cm}
\end{table}

\begin{figure*}[t]
    \vspace{-0cm}
    \centering
    \includegraphics[width =1\linewidth]{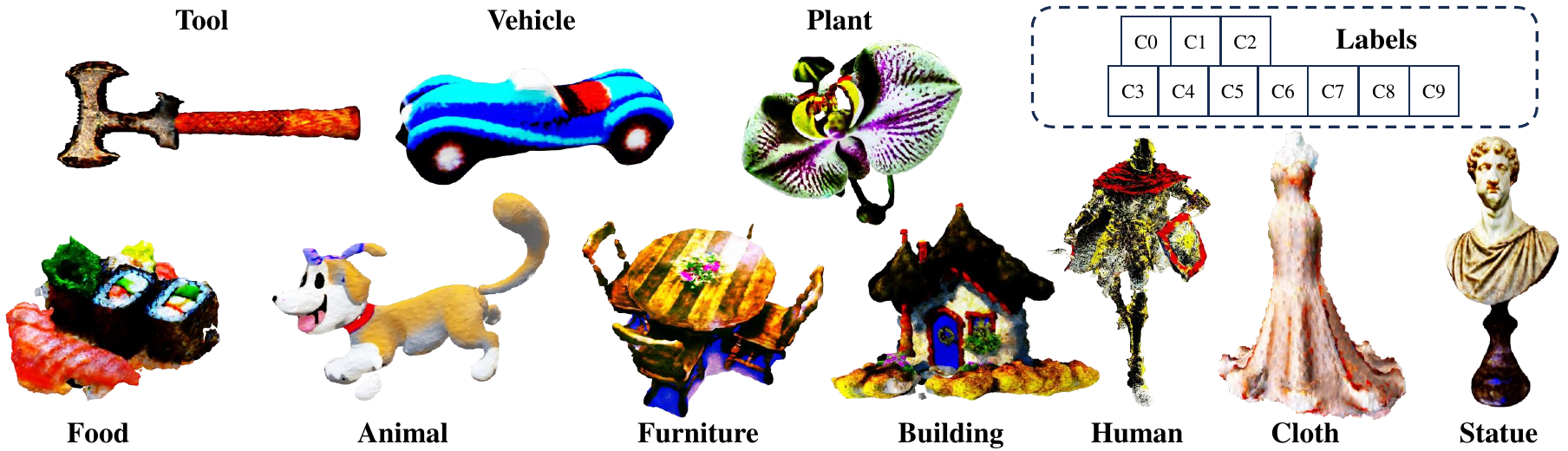}
    \vspace{-0.6cm}
    \caption{Overview of 3DGCs in the 3DGCQA database and corresponding category labels.}
    \label{fig:shot}
    \vspace{-0.4cm}
\end{figure*}

\begin{figure}[t]
    \vspace{-0cm}
    \centering
    \includegraphics[width =1\linewidth]{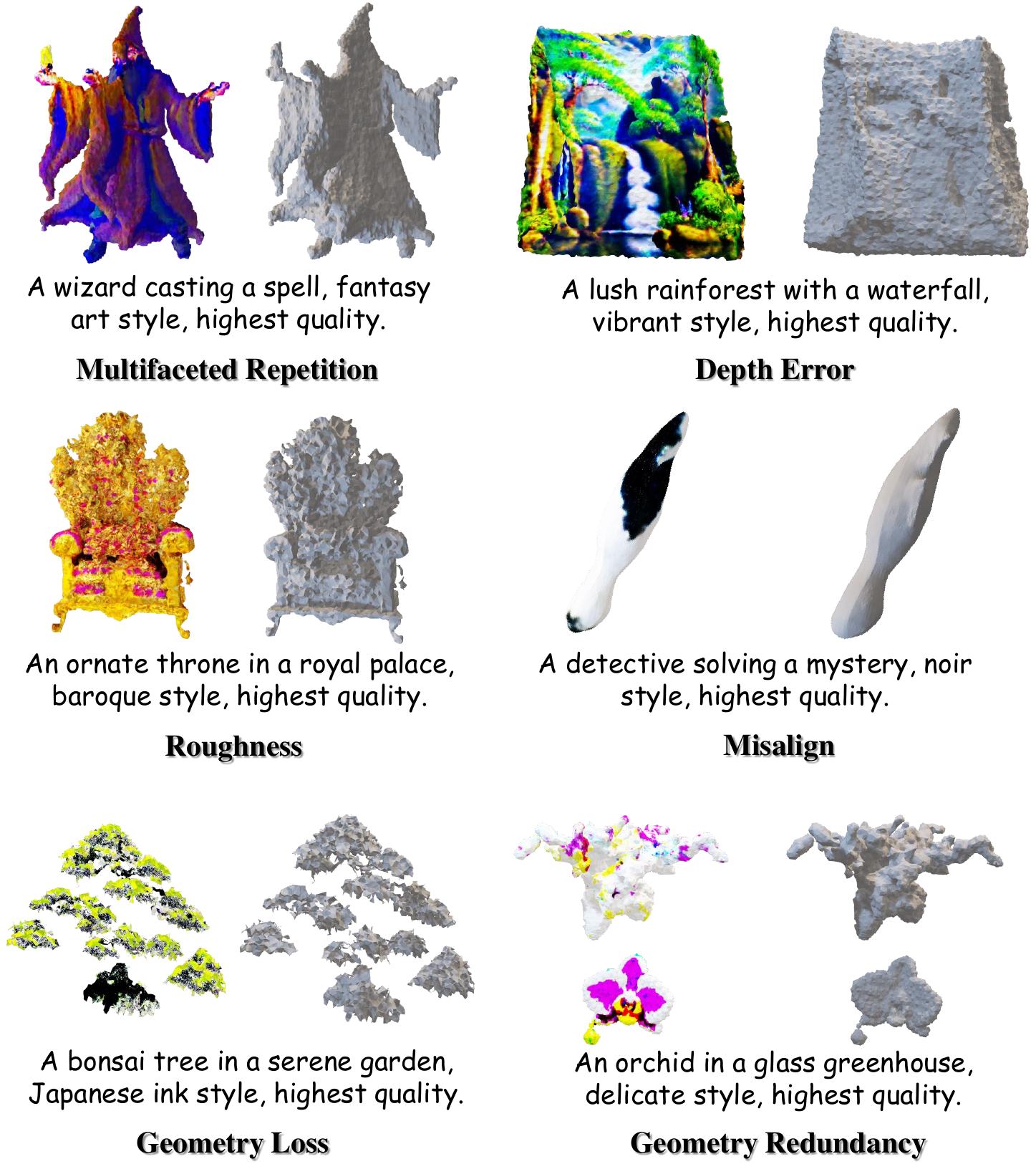}
    \vspace{-0.6cm}
    \caption{Illustration of common distortions occurred in 3DGCs.}
    \label{fig:dis}
    \vspace{-0.4cm}
\end{figure}

\subsection{Prompts}
To ensure the diversity of content in the created dataset, 50 distinct prompts are selected. These prompts are categorized into ten groups, including tools, vehicles, plants, foods, animals, furniture, buildings, human, clothes, and statues. To further illustrate the selected prompts, a word cloud is generated, as shown in Fig.~\ref{fig:prep}(a). Observing the word cloud provides valuable insights into the composition of the dataset. To enhance the quality of the generated 3D content, a \textit{"best quality"} constraint is applied to each prompt. Additionally, the selected prompts incorporate various generation styles, contributing to the dataset’s richness. The predominance of nouns in the word cloud further suggests that the 3DGCQA dataset encompasses a broad range of 3D generated content, reflecting its comprehensive coverage across different categories.

\subsection{Selected Generation Methods}
As shown in Table~\ref{tab:methods}, seven representative Text-to-3D methods are selected to generate 3DGCs from the chosen prompts. It is important to note that these Text-to-3D methods first generate images from the prompts using Text-to-Image (T2I) models, and then further produce 3DGCs based on these images. Consequently, variations in T2I models lead to differences in the quality of the resulting 3DGCs. All selected methods are implemented using Threestudio \cite{threestudio2023}. To evaluate the performance and characteristics of each method, the number of successful and failed generation attempts, as well as the generation time for each method, are recorded, as depicted in Fig.~\ref{fig:prep}(b-c). In Fig.~\ref{fig:prep}(c), some outliers are excluded to provide a more accurate reflection of the time required by each method. Upon analysis, it is found that DreamFusion \cite{poole2022dreamfusion} generated the least diverse content, whereas HiFA \cite{zhu2023hifa}  demonstrates higher flexibility in handling various prompts. However, HiFA's generation time is significantly longer compared to the other methods. Additionally, by comparing DreamFusion and Magic3D \cite{lin2023magic3d}, which employed different T2I models, it is observed that although Stable Diffusion offers faster generation times, it imposes stricter requirements on prompt wording, indicating its limited generalization capability. Overall, the existing Text-to-3D methods exhibit notable limitations in terms of either generalizability or generation speed.

\subsection{Visualization and Distortions}
Fig.~\ref{fig:shot} presents examples of the generated 3DGC content. Upon reviewing Fig.~\ref{fig:shot} , it can be concluded that the selected generation methods are generally effective in producing the required 3DGCs. However, the quality of the \textit{"Human"} model, as depicted, is notably inferior compared to other generated contents. To further investigate potential distortions in 3DGCs, a comprehensive preview of all generated 3DGCs is conducted. This review identifies six common distortions, which are summarized in Fig.~\ref{fig:dis}. Each of these distortions exhibits distinct characteristics and highlights areas for improvement, providing valuable insights for the future development of 3D generative technologies.

\subsection{Subjective Experiment}
\begin{figure}[t]
    \centering
    \vspace{-0cm}
    
    \subfigure[]{\includegraphics[width=0.475\linewidth]{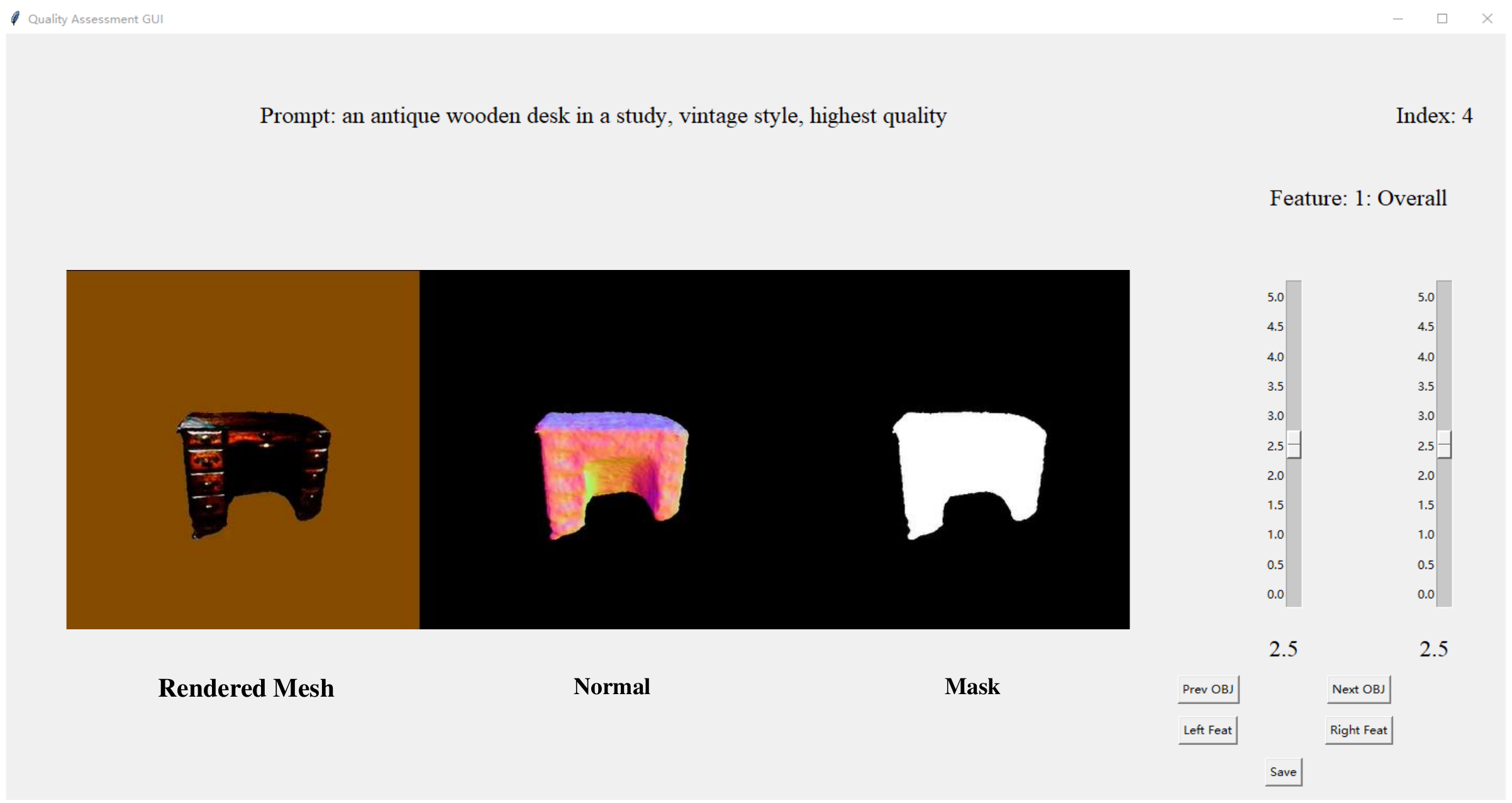}}
    \subfigure[]{\includegraphics[width=0.475\linewidth]{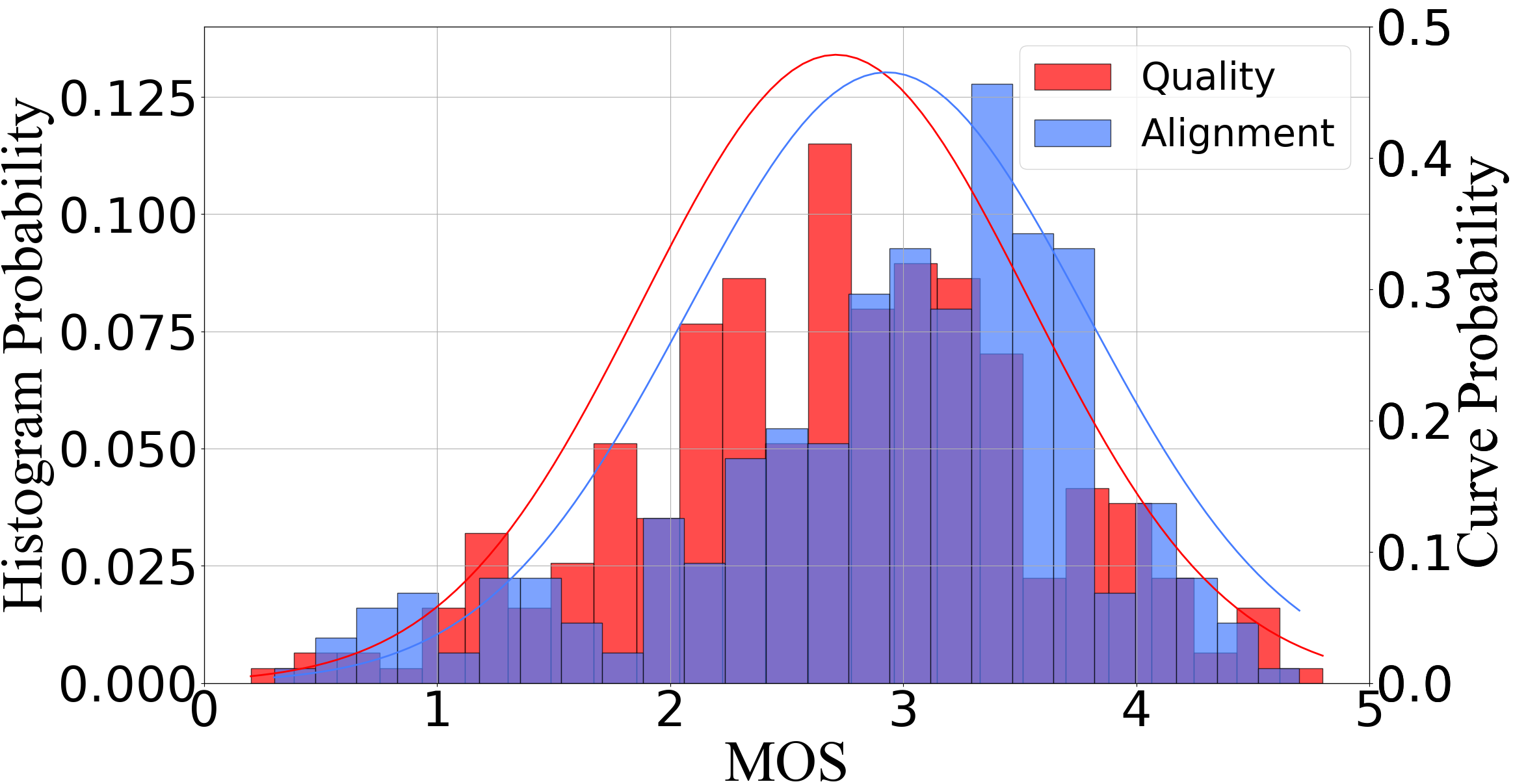}}
    
    \subfigure[]{\includegraphics[width=0.475\linewidth]{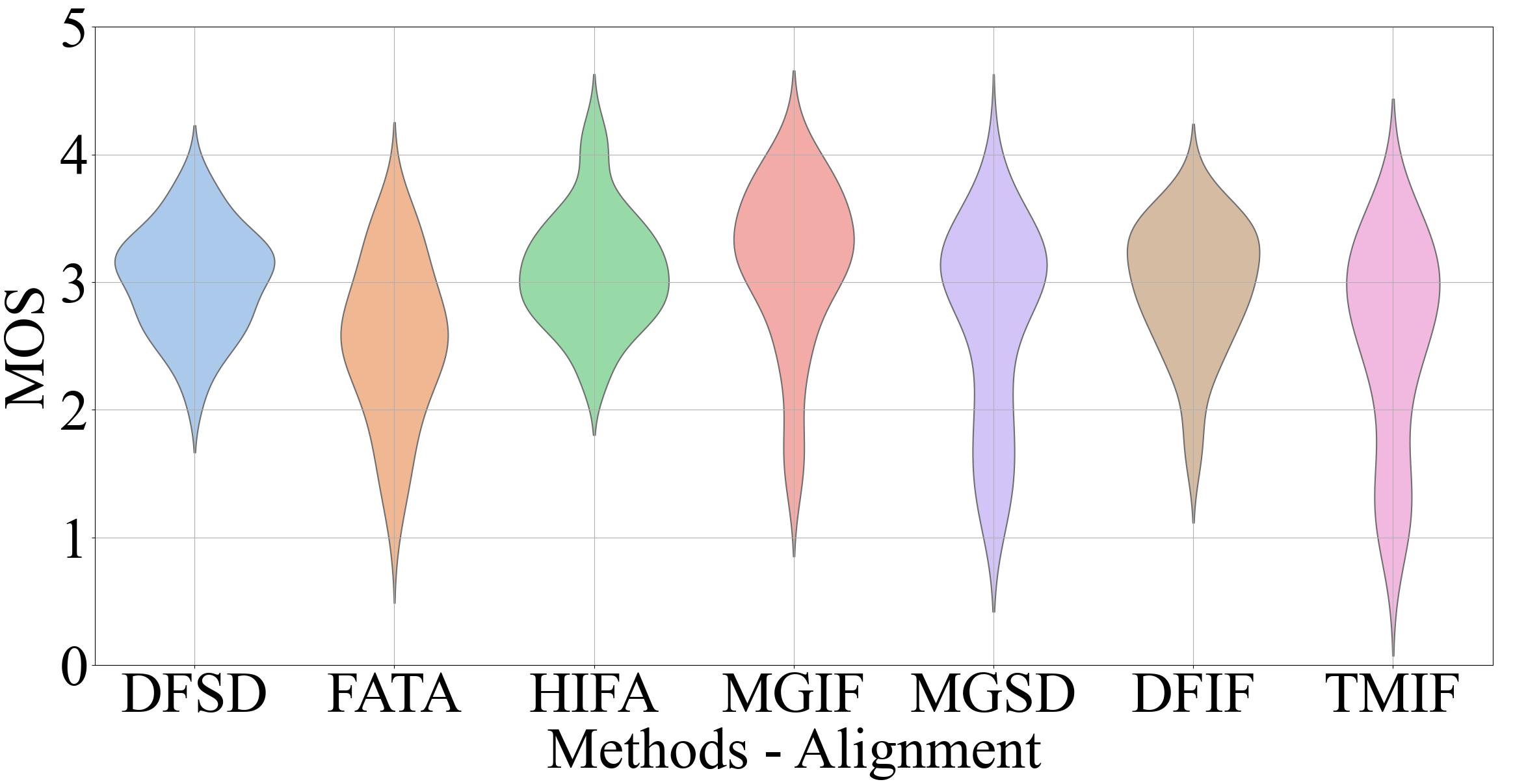}}
    \subfigure[]{\includegraphics[width=0.475\linewidth]{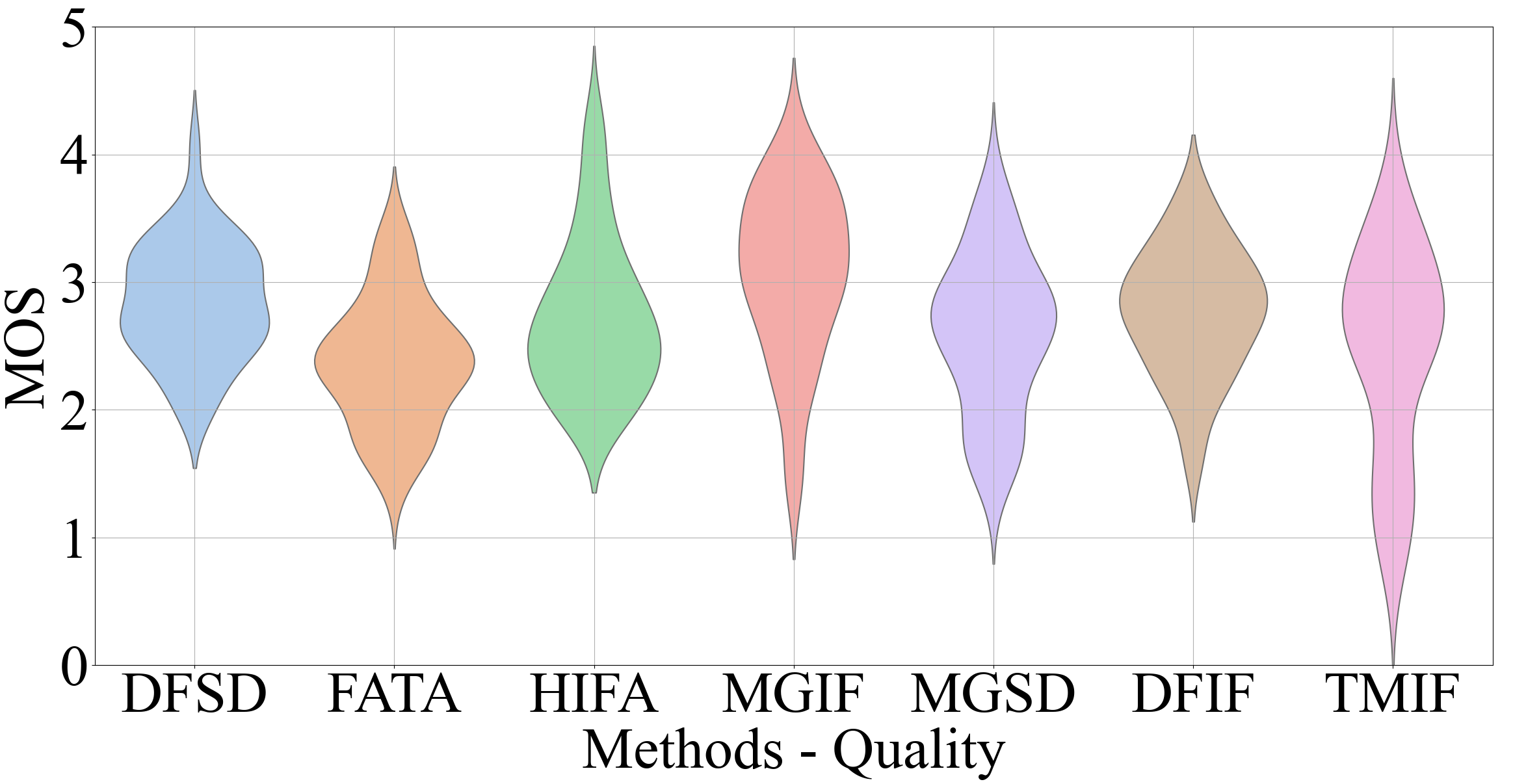}}

    \subfigure[]{\includegraphics[width=0.475\linewidth]{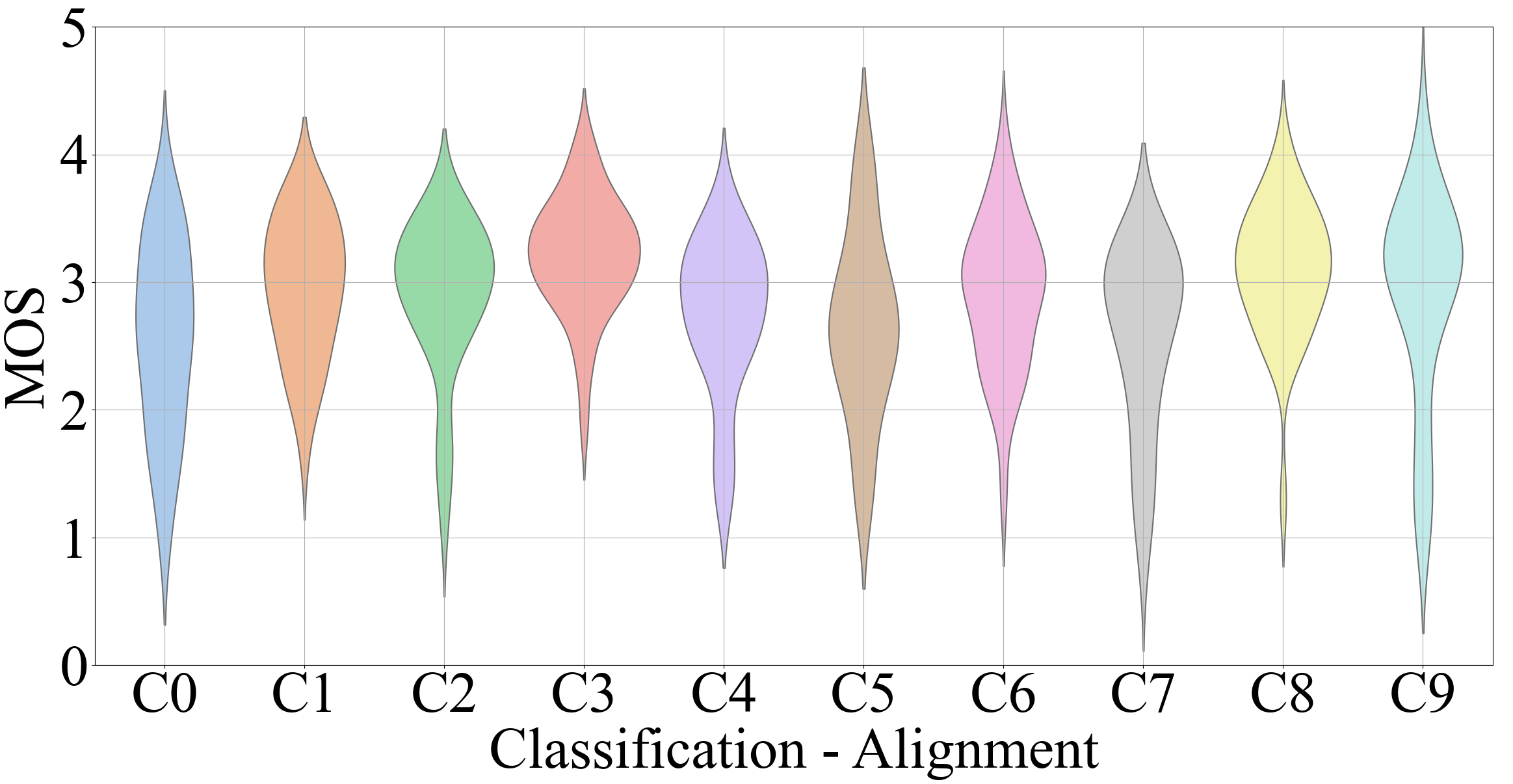}}
    \subfigure[]{\includegraphics[width=0.475\linewidth]{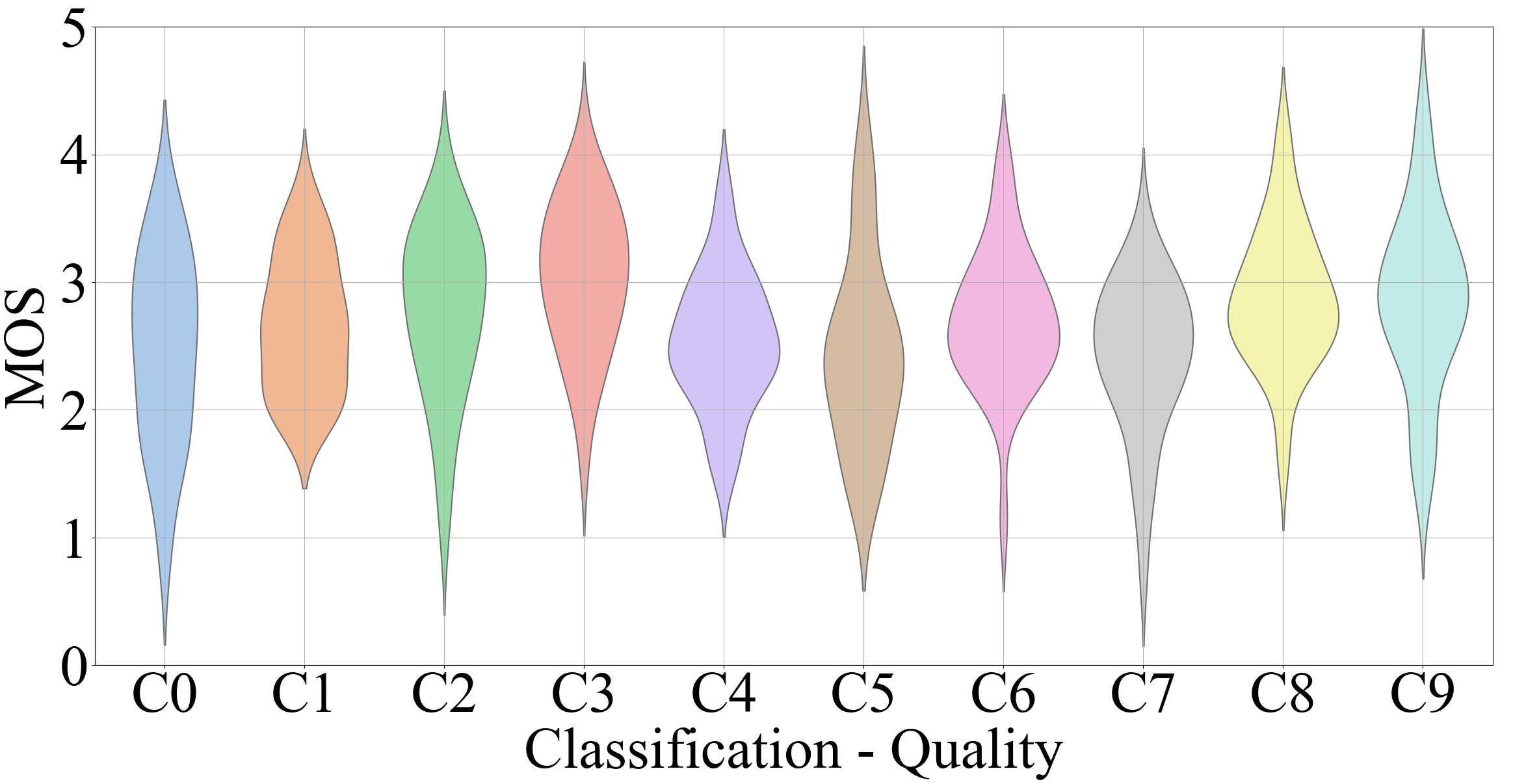}}
    
    \vspace{-0.2cm}
    \caption{The evaluation interface used and distribution of subjective scores.}
    \label{fig:distribution}
    \vspace{-0.4cm}
\end{figure}

To further evaluate the quality of 3D Generated Content (3DGC), subjective scoring experiments are conducted using the established 3DGCQA dataset. Prior to the experiment, all 3DGCs are rendered as videos, with normal and depth information provided for reference. In accordance with the recommendations outlined in ITU-R BT.500-13 \cite{bt2002methodology}, 20 male and 20 female participants are invited to a laboratory setting to subjectively score the rendered 3DGCs. The evaluations are performed using an iMac monitor supporting a resolution of 4096×2304 pixels. The experimental design employs absolute category ratings to assess two key aspects: the consistency of 3DGCs with the given prompts, and the overall quality of the 3DGCs. The 313 3DGCs are divided into six groups: the first five groups contain 50 3DGCs each, while the remaining group contains the final 63 3DGCs. To ensure the reliability of the subjective assessments and minimize potential issues related to visual fatigue or dizziness from prolonged exposure to 3D contents, participants are limited to a maximum of two groups per day, with at least a 30-minute break between groups. Additionally, all participants undergo a 30-minute training session on the first day of the experiment, during which they are introduced to the scoring interface and familiarized with the scoring criteria, as illustrated in Fig.~\ref{fig:distribution}(a). This training ensures participants are adequately prepared for the evaluation process.

At the end of the experiment, a total of 12,520 = 313 × 40 subjective ratings are collected. Each subjective score can be represented as a binary pair ($a_{ij}$,$q_{ij}$), where $a_{ij}$ indicates the alignment score and $q_{ij}$ represents the quality score given by the $i$-th subject for the $j$-th 3DGC. Further, the z-score for each rating can be calculated using the following formula:
\begin{equation}
z_{ij}^s = \frac{{{s_{ij}} - \mu _i^s}}{{\sigma _i^s}}(s \in \{ a,q\} ),
\end{equation}
\noindent where $s_{ij}$ represents the score provided by the $i$-th subject on the $j$-th 3DGC, $\mu^{s}_{i}=\frac{1}{N_{i}} \sum_{j=1}^{N_{i}} s_{i j}$, $\sigma^{s}_{i}=\sqrt{\frac{1}{N_{i}-1} \sum_{j=1}^{N_{i}}\left(s_{i j}-\mu^{s}_{i}\right)}$, and $N_i$ represents the total number of 3DGCs evaluated by subject $i$ .Following the subject rejection procedure recommended in \cite{bt2002methodology}, ratings from unreliable subjects are excluded. The remaining z-scores for alignment $z_{ij}^a$ and quality $z_{ij}^q$ are linearly rescaled to the range [0,5]. Finally, the mean opinion scores (MOSs) for the  $j$-th 3DGC are computed by averaging the rescaled z-scores.

\subsection{Subjective Experiment Analysis}
Based on the results of the subjective experiment, the subjective opinion scores for the alignment and quality dimensions follow the distribution shown in Fig.~\ref{fig:distribution}(b). As can be seen in Fig.~\ref{fig:distribution}(b), although most of the 3DGCs are able to align with the prompts, the quality of their generation still stays at around 2.5 points, indicating that there is indeed a general quality problem in 3DGCs, which re-emphasizes the necessity of conducting quality assessment. To further explore the effects of different methods and different prompt categories on MOS, Fig.~\ref{fig:distribution}(c-f) are plotted. By looking at these figures, some conclusions can be drawn. 1) Analyzing the quality aspects of the generation methods, Magic3D \cite{lin2023magic3d} achieves optimal performance. In addition, DeepFolyd IF \cite{saharia2022photorealistic} is more capable of generating high-quality 3DGCs compared to Stable Diffusion \cite{rombach2022high}. 2) Analyzing from the aspect of prompts, the existing methods produce the lowest quality for \textit{"human,"} which is mainly due to the fact that 3D characters have more complex texture details and strict geometrical structures. 3) In general, various generation methods and various classes of prompts have result in more significant quality differences.

\begin{table}[!t]\centering
\caption{Performance comparison of different methods on 3DGCQA database. The best performance results are marked in \textbf{BOLD}.}
\label{tab1}
\setlength{\tabcolsep}{1.5mm}
\begin{tabular}{c|c|c|c|c|c}
\toprule
Type & Method &  SRCC$\uparrow$ & PLCC$\uparrow$ & KRCC$\uparrow$ & RMSE$\downarrow$\\
\hline
\multirow{6}{40pt}{\centering Hand-craft Features}
& BRISQUE \cite{brisque} & 0.2091 & 0.3347 & 0.1444 & 0.7414\\
& CPBD \cite{narvekar2011no} & 0.2099 & 0.4797 & 0.1335 & 0.7217\\
& IL-NIQE \cite{ilniqe} & 0.1481 & 0.1573 & 0.1131 & 0.6600\\
& NFERM \cite{nferm}& 0.2797 & 0.4062 & 0.1999 & 0.7222\\
& NFSDM \cite{nfsdm}& 0.3189 & 0.4468 & 0.2235 & 0.6935\\
& NIQE \cite{niqe} & 0.2079 & 0.2594 &0.1413 & 0.8050\\ \hdashline
\multirow{2}{40pt}{\centering Deep Learning}
& DBCNN \cite{dbcnn}& \bf{0.5381} & \bf{0.5147} &\bf{0.3946} &  \bf{0.4700}\\
& StairIQA \cite{stairiqa}& 0.3813 & 0.4566 & 0.2653 &0.5802\\
\bottomrule
\end{tabular}
\vspace{-0.6cm}
\end{table}

\section{Benchmark Experiment}
\subsection{Experimental Setup}
For benchmarking the performance of the proposed 3DGCQA dataset, several standard quality assessment methods are selected to provide a reference for future algorithm development. Specifically, eight projection-based assessment methods are included in the experiment. Among these, DBCNN and StairIQA are deep learning-based approaches, while the remaining methods rely on feature extraction for quality assessment. It is worth stating that all evaluation algorithms are implemented using source code provided by the authors. The 3DGCQA dataset is randomly divided into five folds, ensuring no overlap in content between the folds. To evaluate performance, a five-fold cross-validation approach is employed, and the average results across all folds are reported. The performance of each algorithm is measured using four widely used metrics: Spearman rank order correlation coefficient (SRCC), Pearson linear correlation coefficient (PLCC), Kendall rank order correlation coefficient (KRCC), and root mean square error (RMSE).

\subsection{Performance \& Analysis}
The experimental results are presented in Table~\ref{tab1}, which demonstrate that current quality assessment methods are not well-suited for evaluating the quality of 3DGCs. For these methods, the ineffectiveness may stem from the loss of geometric and depth information, which obscures certain quality-related issues. These findings highlight the need for the development of novel assessment frameworks specifically designed to address the unique challenges posed by 3DGCs.

\section{Conclusion}
As generative artificial intelligence (AI) continue to advance, increasing attention is being directed towards the quality of generated content. In the case of 3D generative content (3DGC), quality issues are particularly pronounced due to the complexity of its data structure. To address these challenges, this paper introduces the 3DGCQA dataset, a quality assessment dataset specifically for 3DGCs. The dataset consists of 313 textured meshes generated by seven representative Text-to-3D methods, based on ten distinct categories of prompts. Through multi-dimensional analysis of the selected prompts, generation methods, and results from subjective experiments, it is evident that current 3D generation techniques exhibit limitations in generalization, generation time, and quality. These findings underscore the need for robust quality assessment mechanisms. Finally, the paper evaluates several existing quality assessment methods on the 3DGCQA dataset, providing valuable benchmarks and references for the future development of new assessment methodologies.

\bibliographystyle{IEEEtran}
\bibliography{IEEEexample}

\end{document}